%% file: main.tex
 \documentclass[acmsmall,screen]{acmart}
 
\usepackage{amsmath,amsfonts}
\usepackage{textcomp}
\usepackage{algorithm}
\usepackage[noend]{algpseudocode}
\usepackage{amsmath}
\usepackage{enumitem}
\usepackage{rotating}
\usepackage{graphicx}
\usepackage{multirow}
\usepackage{xspace}
\usepackage[normalem]{ulem}
\usepackage{color}
\usepackage{pifont}
\usepackage{bbding}
\usepackage{svg}
\usepackage{threeparttable}
\usepackage{booktabs}
\usepackage{balance}
\usepackage{tabularx}
\usepackage{bm}
\usepackage{xcolor}
\usepackage{colortbl}
\usepackage{color}
\usepackage{transparent}
\usepackage{import}
\usepackage{listings}
\usepackage{subcaption}
\usepackage{hyperref}
\usepackage{cleveref}
\usepackage{verbatim}
\usepackage[many]{tcolorbox}
\usepackage{flushend}

\newcommand{\lida}[1]{\textcolor{black}{#1}}

\newcommand{\toolname}{ECScan\xspace}

\usepackage{enumitem}

\newtcolorbox{titleEnv}{
colframe=black!80,
colback=gray!10,
fonttitle=\bfseries,
coltitle=black,
left=3pt,
right=3pt,
top=3pt,
bottom=3pt,
boxrule=0.4mm,
arc=3mm
}

\definecolor{codegreen}{rgb}{0,0.6,0}
\definecolor{codegray}{rgb}{0.5,0.5,0.5}
\definecolor{codepurple}{rgb}{0.58,0,0.82}
\definecolor{backcolour}{rgb}{0.95,0.95,0.92}

\lstdefinestyle{mystyle}{
    commentstyle=\color{codegreen},
    keywordstyle=\color{magenta},
    numberstyle=\tiny\color{codegray},
    stringstyle=\color{codepurple},
    basicstyle=\ttfamily\footnotesize,
    breakatwhitespace=false,         
    breaklines=true,                 
    captionpos=b,                    
    keepspaces=true,                 
    numbers=left,                    
    numbersep=2pt,                  
    showspaces=false,                
    showstringspaces=false,
    showtabs=false,                  
    tabsize=1
}

\begin{document}
\title{Detecting Essence Code Clones via Information Theoretic Analysis}

\input{authors}

\thispagestyle{plain}
\pagestyle{plain}
\input{abstract}
\maketitle
\input{content}

\balance
\bibliographystyle{ACM-Reference-Format}
\bibliography{ref}

\end{document}

%% file: authors.tex
\author{Lida Zhao}
\affiliation{%
  \institution{Nanyang Technological University}
  \country{Singapore}
}
\email{lida001@e.ntu.edu.sg}

\author{Shihan Dou}
\affiliation{%
  \institution{Fudan University}
  \country{China}
}
\email{shihandou@foxmail.com}

\author{Yutao Hu}
\affiliation{%
  \institution{Huazhong University of Science and Technology}
  \country{China}
}
\email{yutaohu@hust.edu.cn}

\author{Yueming Wu}
\affiliation{%
  \institution{Huazhong University of Science and Technology}
  \country{China}
}
\email{wuyueming21@gmail.com}

\author{Jiahui Wu}
\affiliation{%
  \institution{Nanyang Technological University}
  \country{Singapore}
}
\email{jiahui004@e.ntu.edu.sg}

\author{Chengwei Liu}
\affiliation{%
  \institution{Nanyang Technological University}
  \country{Singapore}
}
\email{chengwei.liu@ntu.edu.sg}

\author{Lyuye Zhang}
\affiliation{%
  \institution{Nanyang Technological University}
  \country{Singapore}
}
\email{zh0004ye@e.ntu.edu.sg}

\author{Yi Liu}
\affiliation{%
  \institution{Nanyang Technological University}
  \country{Singapore}
}
\email{yi009@e.ntu.edu.sg}

\author{Jun Sun}
\affiliation{%
  \institution{Singapore Management University}
  \country{Singapore}
}
\email{junsun@smu.edu.sg}

\author{Xuanjing Huang}
\affiliation{%
  \institution{Fudan University}
  \country{China}
}
\email{xjhuang@fudan.edu.cn}

\author{Yang Liu}
\affiliation{%
  \institution{Nanyang Technological University}
  \country{Singapore}
}
\email{yangliu@ntu.edu.sg}

\authorsaddresses{%
Authors’ addresses: L. Zhao, J. Wu, C. Liu, L. Zhang, Y. Liu, and Y. Liu are with Nanyang Technological University, Singapore; emails: \{lida001, jiahui004@, zh0004ye, yi009\}@e.ntu.edu.sg, \{chengwei.liu, yang.liu\}@ntu.edu.sg.
S. Dou and X. Huang are with Fudan University, China; emails: shihandou@foxmail.com, xjhuang@fudan.edu.cn.
Y. Hu and Y. Wu (corresponding author) are with Huazhong University of Science and Technology, China; emails: wuyueming21@gmail.com, yutaohu@hust.edu.cn.
}

%% file: abstract.tex
\begin{abstract}
Code cloning, a widespread practice in software development, involves replicating code fragments to save time but often at the expense of software maintainability and quality. In this paper, we address the specific challenge of detecting ``essence clones'', a complex subtype of Type-3 clones characterized by sharing critical logic despite different peripheral codes. Traditional techniques often fail to detect essence clones due to their syntactic focus. To overcome this limitation, we introduce \toolname, a novel detection tool that leverages information theory to assess the semantic importance of code lines. By assigning weights to each line based on its information content, \toolname emphasizes core logic over peripheral code differences. Our comprehensive evaluation across various real-world projects shows that \toolname significantly outperforms existing tools in detecting essence clones, achieving an average F1-score of 85\%. It demonstrates robust performance across all clone types and offers exceptional scalability. This study advances clone detection by providing a practical tool for developers to enhance code quality and reduce maintenance burdens, emphasizing the semantic aspects of code through an innovative information-theoretic approach.
\end{abstract}




%% file: content.tex
\section{Introduction}



\begin{figure*}[htbp]
\begin{minipage}[t]{0.5\textwidth}
\begin{lstlisting}[escapechar=@, language=Java]
public void put(String key, InputStream in) {
    File f = new File(mStoreDir, key);
    OutputStream os = new BufferedOutputStream(new FileOutputStream(f));
    try {
        @\colorbox{yellow}{byte[] data = new byte[1024];}@
        @\colorbox{yellow}{int nbread;}@
        @\colorbox{yellow}{while ((nbread = in.read(data)) != -1)}@
        @\colorbox{yellow}{os.write(data, 0, nbread);}@
    } finally {
        @\colorbox{yellow}{in.close();}@
        @\colorbox{yellow}{os.close();}@
    }
}
\end{lstlisting}
\end{minipage}%
\hfill 
\begin{minipage}[t]{0.5\textwidth}
\begin{lstlisting}[escapechar=@, language=Java]
public void doGet(HttpServletRequest req, HttpServletResponse res){
    res.setContentType("text/html");
    ServletOutputStream out = res.getOutputStream();
    InputStream in = null;
    try {
        URL url = new URL(req.getParameter("loc"));
        in = url.openStream();
        @\colorbox{yellow}{byte[] buffer = new byte[4096];}@
        @\colorbox{yellow}{int bytes\_read;}@
        @\colorbox{yellow}{while ((bytes\_read = in.read(buffer)) != -1)}@
        @\colorbox{yellow}{out.write(buffer, 0, bytes\_read);}@
    } catch (MalformedURLException e) {
        System.err.println(e.toString());
    } finally {
        @\colorbox{yellow}{in.close();}@
        @\colorbox{yellow}{out.close();}@
    }
}
\end{lstlisting}
\end{minipage}
\caption{Motivation Example. The core logic is highlighted in yellow.}
\label{fig:motivation}
\end{figure*}

Code cloning, defined as the replication of code fragments \cite{ain2019systematic}, encompasses both syntactic and semantic code clones. 
Syntactic code clones, including Type-1, Type-2, and Type-3 clones, exhibit similarities in structure. 
Conversely, semantic code clones (i.e., Type-4) achieve analogous functionality using different syntactic forms. 
Despite its potential time-saving benefits for developers, code cloning presents numerous drawbacks. 
Studies have shown that it can lead to a decline in overall code quality \cite{kim2005empirical, lavoie2010challenging}, escalate software maintenance costs \cite{mayrand1996experiment}, and heighten the risk of vulnerability propagation \cite{baker1995finding, pham2010detection, li2016clorifi}. 
Moreover, legal disputes may arise due to copyright infringement issues associated with cloned code \cite{xu2023lidetector}. 
Given these challenges, there is a growing emphasis on developing effective clone detection techniques within the software engineering domain.

In real-world development scenarios, developers tend to copy code snippets that implement core logic or algorithms because these snippets are more valuable and provide greater convenience. 
As depicted in \Cref{fig:motivation}, the function on the left stores data from an input stream into a file, while the function on the right reads data from a URL specified by the client and writes it back to the client's response. Despite differing in function names, parameters, and types of input and output streams, the core logic, highlighted in yellow, shares the same principle of transferring data using buffered I/O operations within a while loop. 
For this type of clone, a part of the code is similar or identical across two functions, representing the core logic or most crucial aspect of those functions. 
Even if the surrounding code differs, we still consider them as a clone pair since the essence or foundational logic of the functions is the same. 
We refer to this type of clone as ``essence clone'', which falls under the complex Type-3 clones. 
Given that this scenario also occurs in real development environments, there is an urgent need to develop a tool that can support large-scale essence clone detection.

Nowadays, numerous methods for code clone detection have been proposed, with some aiming to enhance the detection of semantic clones by extracting intermediate representations (e.g., abstract syntax trees and program dependency graphs) of the code.
Tree-based approaches \cite{jiang2007deckard, wei2017cdlh, zhang2019astnn, liang2021astpath, jo2021twopass} extract the parse tree of a program to maintain the syntax of the program.
Graph-based approaches \cite{krinke2001duplix, komondoor2001pdgdup, wang2017ccsharp, zhao2018deepsim, zou2020ccgraph} extract the graph structure of the code, containing more semantic details of the code and enabling more effective detection of semantic clones.
Traditionally, both methods use tree matching or graph mining for clone detection, and they often have high computational costs and lack scalability with large datasets.
In fact, existing studies have demonstrated that the majority of clones in software systems belong to Type-3 clones~\cite{roycordy2010,svajlenko2014big}. 
Therefore, to meet real-world needs, more attention should be paid to large-scale Type-3 clone detection.

Existing token-based techniques \cite{gode2009incremental, kamiya2002ccfinder, sajnani2016sourcerercc, li2017cclearner, wang2018ccaligner, golubev2021multi, hung2020cppcd} convert the code into a sequence of tokens and then analyze token similarity for large-scale Type-3 code clone detection.
For instance, \emph{SourcererCC} \cite{sajnani2016sourcerercc} detects Type-3 code clones by comparing the number of overlapping tokens between two code snippets and can scale to scan code clones among 428 million files \cite{lopes2017dejavu}. 
However, it is challenging to detect essence clones when the essence code occupies only a small part of the entire function from the perspective of token count.
Referring to the example in \Cref{fig:motivation}, due to the low proportion of core logic tokens and variations in variable names, such clones are not detected by traditional clone detection tools. 
Specifically, the ratios of tokens involved in the core logic to the total number of tokens in the two functions are 0.60 and 0.40, respectively. 
Using SourcererCC~\cite{sajnani2016sourcerercc}, the similarity between these functions is approximately 0.3, significantly below the threshold of 0.7 set in their original study.

In order to facilitate the detection of large-scale Type-3 code clones, particularly essence clones, we have developed a novel tool, named \textbf{E}ssense \textbf{C}lone \textbf{Scan}ner (\toolname), specifically designed to enhance the detection of such clones. 
The tool assigns weights to expressions based on the significance of the code logic, emphasizing the principle that more critical logic should carry greater weight. 
Assessing semantic importance directly is challenging, but as demonstrated by Wu et al.~\cite{ossfp}, importance is positively correlated with the information content of a code block. 
We then assign weights to code lines based on their information content. 
Inspired by Aizawa et al.~\cite{AIZAWA200345}, we first calculate the ``probability-weighted amount of information'' for each token using Term Frequency-Inverse Document Frequency (TF-IDF), and then aggregate these values to determine the information content of each line. 
The weight of each line is then set as its proportion of the total information score. 
During the process of identifying potential clone candidates, we generate N-line hashes as features for each code block and construct an inverse hash index for these features. 
Code blocks that share the same N-line features are grouped together, pending further verification.
For verification, we calculate the similarity of potential clone pairs by treating each line as a unique dimension and setting the weight of the line as the value of that dimension.
We then use cosine similarity to evaluate the similarity between clone pairs. 
Consequently, clones that include more important lines of code will yield a higher similarity score, even if they constitute a relatively small proportion of the total token count.

To evaluate the performance and efficiency of \toolname, we first tested \toolname on the BigCloneBench~\cite{big} and compared it with 7 other tools, including CCAligner~\cite{wang2018ccaligner}, SourcererCC~\cite{sajnani2016sourcerercc}, Siames~\cite{Ragkhitwetsagul2019SiameseSA}, NIL~\cite{nil2021}, Nicad~\cite{roy2008nicad}, LVMapper~\cite{lvmapper2020}, and Decard~\cite{jiang2007deckard}. 
The results indicate that \toolname performs well in detecting all Type-1, Type-2, achieving f1-score of 99\%, 98\%, and have top-rank performance in 4 subtypes of Type-3 clones. 
We then conducted experiments on four carefully selected real-world Java projects and compared them with two state-of-the-art tools, NIL~\cite{nil2021} and SourcererCC~\cite{sajnani2016sourcerercc}. 
The results show that \toolname consistently outperforms both tools across various projects, demonstrating its enhanced capability to accurately detect essence code clones while maintaining a high detection rate of true clones. 
The average F1-score achieved by \toolname is 85\%, compared to 70\% and 61\% for NIL and SourcererCC, respectively.
Additionally, we examined the scalability of \toolname using the IJaDataset~\cite{ijadataset} with data sizes ranging from 1K to 10M, where \toolname exhibited moderate to high scalability.

The main contributions of this paper are as follows
\begin{itemize}
    \item \lida{We identify a type of Type-3 clone, often overlooked by traditional clone detection techniques, called an ``essence clone''. To address this, we propose an innovative method that assigns weights to lines of code based on the information they carry, enabling more effective detection.}
    \item We have successfully implemented this technique in a software tool named \toolname. This integration allows for practical application and usability in real-world programming environments. 
    \item Our evaluations of \toolname show that it excels at identifying essence clones more effectively than current methods, and performs well against the top clone detectors for general Type-1, Type-2, and Type-3 clone detection.
\end{itemize}

\noindent \textbf{Paper organization.} 
The rest of the paper is organized as follows: 
Section 2 presents a comprehensive discussion on the different types of code clones, providing a detailed definition and classification to clarify the context of our study.
Section 3 provides a detailed description of our system, outlining the architecture and implementation details of \toolname for detecting essence clones.
Section 4 reports the evaluation results, where we systematically analyze the performance of \toolname against a well-known benchmark and four real-life projects and other leading tools in the field. We provide metrics such as precision, recall, and F1-score to demonstrate its efficacy.
Section 5 discusses the paper about the relationships between essence clone and other types of clones.
Section 6 describes the related work. 
Section 7 concludes the paper.

\section{Preliminaries}

\subsection{Clone Definition}
A code snippet is a consecutive segment of source code identifiable by its file name and the starting and ending line numbers. A code block within this context is a code snippet enclosed by braces, and for our study, we define a function as the code block, consistent with previous studies~\cite{sajnani2016sourcerercc,wang2018ccaligner}. Clones are pairs or groups of code snippets that are either identical or very similar. The smallest recognizable clone spans at least six lines or 50 tokens~\cite{bellon2007type1_4}, which establishes a threshold for identifying significant code repetitions. Each pair of similar code snippets is categorized as a clone pair, each of which is associated with a specific clone type.

Previous works have defined 4 types of clones. 
\begin{itemize}
    \item \textbf{Type-1 Clones} refer to code snippets that are textually similar and are identical except for differences in spaces, blank lines, and comments.
    \item \textbf{Type-2 Clones}, also known as Lexical Similarity, consists of identical code snippets except for variations in identifiers such as function names, class names, and variables.
    \item \textbf{Type-3 Clones}, also known as Syntactic Similarity, are code snippets that differ at the statement level, including additions, modifications, or deletions of statements. 
    \item \textbf{Type-4 Clones} are code snippets that share Semantic Similarities. These snippets may be syntactically different but perform the same function, thus demonstrating functional similarity, as opposed to the textual similarity observed in Type-1, 2, and 3 clones.
\end{itemize}

To address the ambiguity between Type-3 and Type-4 clone pairs, Svajlenko et al. \cite{svajlenko2014big} introduced further granularity by subdividing them into four subtypes based on line-level and token-level similarity: 
\begin{itemize}
    \item \emph{Very Strongly Type-3} (VST3) clones with a similarity range of [0.9, 1.0)
    \item \emph{Strongly Type-3} (ST3) clones with a similarity range of [0.7, 0.9)
    \item \emph{Moderately Type-3} (MT3) clones with a similarity range of [0.5, 0.7) 
    \item \emph{Weakly Type-3/Type-4} (WT3/T4) clones with a similarity range of [0.0, 0.5)
\end{itemize}
In this work, all of the discussion and experiments are based on the classification proposed by Svajlenko et al.~\cite{svajlenko2014big}

\subsection{Essence Clone Definition}\label{sec:essence_def}
An essence clone is defined as a relationship between two code blocks where only a portion of the code needs to be similar, and this similar portion encapsulates the core logic or the most critical aspect of the code block.
Despite differences in the surrounding code, the similar essence or foundational logic of these functions categorizes them as clones of each other. 
This definition requires us to identify which part constitutes the core logic and to what extent each line contributes to the overall logic compared to others. 
Identifying the semantic information within a code block has been a consistent challenge since the emergence of Type 4 clones, and directly quantifying the semantic significance of individual lines is equally challenging. 
In the study by Wu et al.~\cite{ossfp}, it was highlighted that the richness of information embedded within code often corresponds to the significance of its underlying logic. 
Building upon this observation, we introduce an innovative method to tackle the aforementioned challenges through the lens of information theory.
By quantifying the information content within each line of code, we can gauge the richness of core logic details.

Let $ C_k $ be a code block, where each code block is represented as a set of lines $ C_k = \{l_1^{(k)}, l_2^{(k)}, \ldots, l_n^{(k)}\} $.
For each line $ l_i^{(k)} $ in a code block, define:
\begin{itemize}
    \item $ H_{\text{line}}(l_i^{(k)}) $ as the amount of information of line $ l_i^{(k)} $.
    \item $ H_{\text{block}}^{(k)}$ as the total amount of information of the code block, where $$ 
        H_{\text{block}}^{(k)} = \sum_{l^{(k)} \in C_k} H_{\text{line}}(l^{(k)}) 
        $$
    \item $ W_{\text{line}}(l_i^{(k)}) $ represents the proportion of information in line $ l_i^{(k)} $ relative to the total information in the entire code block. This value also reflects the semantic significance of $ l_i^{(k)} $ within the code block $C_k$.
        $$
        W_{\text{line}}(l_i^{(k)}) = \frac{H_{\text{line}}(l_i^{(k)})}{H_{\text{block}}^{(k)}}
        $$
\end{itemize}

Assume there are two code blocks $ C_1 $ and $ C_2 $. 
According to the definition, an essence clone is characterized if the lines of $ C_1$ and $ C_2 $ capture similar core logic or critical aspects of the code block. In other words, if the amount of information of matched lines surpasses a certain proportion, a \textbf{Similarity Threshold $\theta$},  of the total amount of information of the code block, the pair of code blocks should be regarded as essence clones.


Formally, Let $M$ be the set of lines that represent the matched core logic between $C_1$ and $C_2$.
$$ 
    M = \{ (l_i^{(1)}, l_j^{(2)}) | l_i^{(1)} \approx l_j^{(2)}\}
$$
where $l_i^{(1)} \approx l_j^{(2)}$ indicates that lines $l_i^{(1)}$ and $l_j^{(2)}$ are semantically similar or identical.
$ (C_1, C_2) $ is an essence clone if: $$
    \text{verify} ( \{W_{line}(l_i^{(1)}) | (l_i^{(1)}, l_j^{(2)})\in M\}, \{W_{line}(l_i^{(2)}) | (l_i^{(1)}, l_j^{(2)})\in M\}) > \theta
$$ 
where \textit{verify} is the similarity verification function. Our algorithm begins by converting matched block weights into vectors, followed by calculating the cosine similarity between the two blocks. More algorithm details are shown in \Cref{sec:verify}.

\subsection{Relations of Different Clone Types}

\begin{figure}
    \centering
    \includegraphics[width=0.6\linewidth]{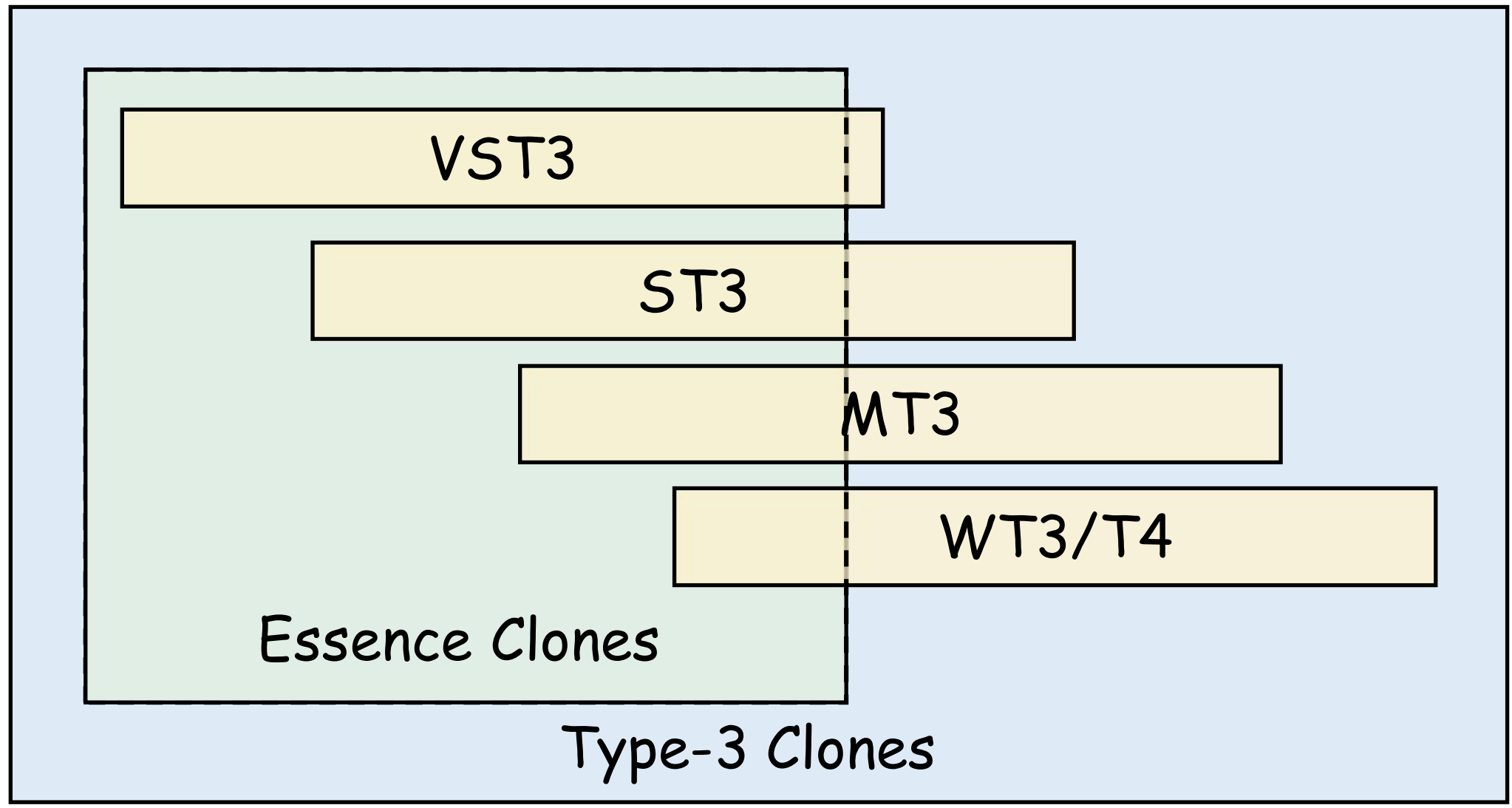}
    \caption{Relations Between Essence Clones and Other Clones}
    \label{fig:clone-relation}
\end{figure}

\Cref{fig:clone-relation} illustrates the relationships between essence clones and four sub-types of Type-3 clones. 

\lida{Type-3 clones, as introduced earlier, are divided into four sub-categories: VST3, ST3, MT3, and WT3/T4. Essence clones exist across all Type-3 subtypes but vary in their distribution. In VST3 and ST3, most code blocks remain identical, often retaining the same core logic, making most instances of VST3 and ST3 recognizable as essence clones. However, as complexity and variability increase, fewer clones from MT3 and WT3/T4 are classified as essence clones. This is because the core logic in MT3 and WT3/T4 may either be inconsistent or differ in ways that make the core elements less comparable. Conversely, an essence clone may not align with MT3 or WT3/T4 if non-core sections take a great proportion or are significantly distinct, causing fewer shared tokens or lines, resulting in lower similarity.}

\section{Methodology}

\begin{figure*}[htbp]
    \centering
    \includegraphics[width=\textwidth]{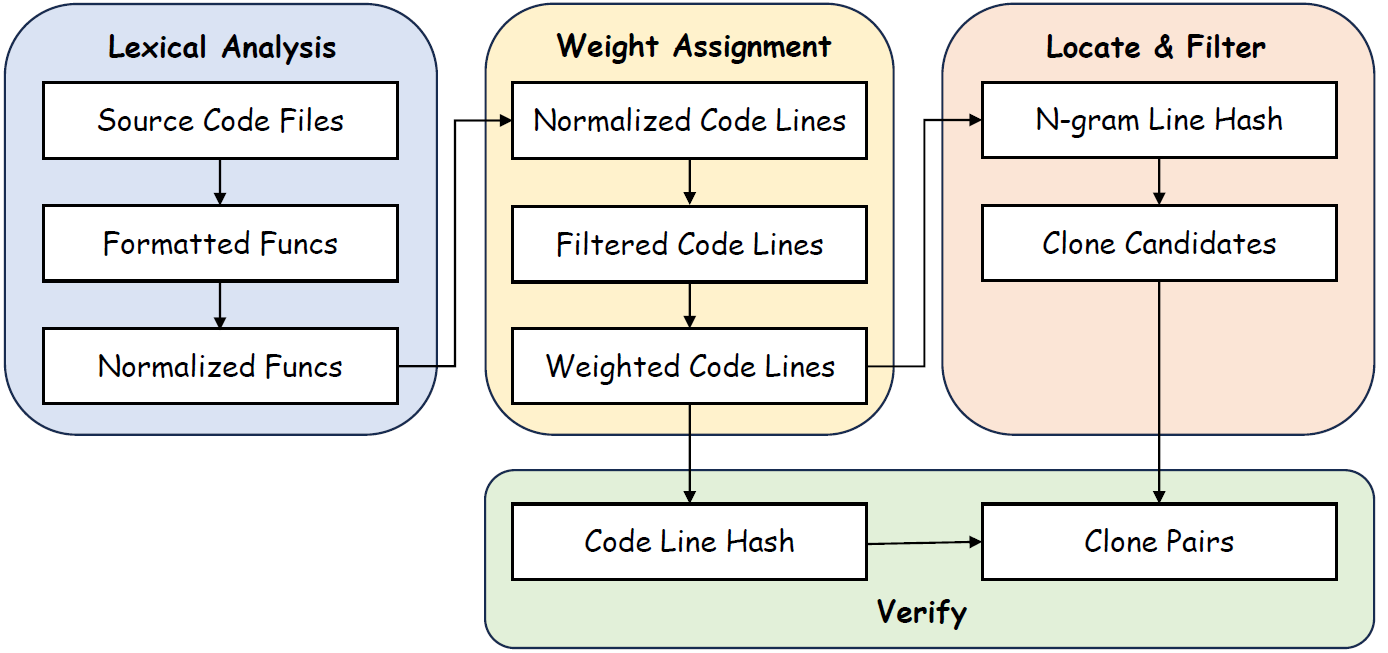}
    \caption{Overview of \toolname}
    \label{fig:overview}
\end{figure*}

In this section, we describe the details of \toolname, including the overview and a detailed description of each phase.

\subsection{Overview}
As shown in Figure \ref{fig:overview}, \toolname consists of four phases: \emph{Lexical Analysis}, \emph{Weight Assignment}, \emph{Locate \& Filter}, and \emph{Verify}.

\begin{itemize}
    \item \textbf{\emph{Lexical Analysis}}: In this phase, blocks of code are extracted from files and subjected to tokenization. These tokens are normalized to account for Type-1 and Type-2 variations.
    \item \textbf{\emph{Weight Assignment}}: In this phase, weights are assigned to each line of code based on its amount of information.
    \item \textbf{\emph{Locate \& Filter}}: In this phase, indexes are constructed for each line to efficiently identify clone candidates.
    \item \textbf{\emph{Verify}}: In this phase, these candidates are verified to produce the final results.
\end{itemize}

\begin{table}[h]
\centering
\caption{Token Types and Their Replacements}
\label{tab:token_replacements}
\begin{tabular}{cc}
\hline
\textbf{Token Type} & \textbf{Replacement} \\ \hline
VariableDeclaratorIdContext & LocVar \\
TypeIdentifierContext & TypeID \\
MethodCallContext & LocMeth \\
QualifiedNameContext & QlfType \\
ClassOrInterfaceTypeContext & ClsType \\
PrimitiveTypeContext & PmtType \\
MethodDeclarationContext & LocMeth \\
LiteralContext & Str \\
IntegerLiteralContext & Num \\
FloatLiteralContext & Num \\
LocalTypeDeclarationContext & LocalType \\
IdentifierContext & ID \\ \hline
\end{tabular}
\end{table}

\subsection{Lexical Analysis}
The objective of Lexical Analysis is to format the code lines and normalize the tokens within each code block. 
This process begins by removing comments and formatting the source code file using the google-java-format~\cite{formatter}. 
This step is crucial because our clone detection technique relies on the individual line of code as the basic unit of detection, and inconsistencies in line breaks or code format can severely affect feature extraction and consequently clone detection accuracy. 
It also eliminates Type-1 variations. 
The google-java-format is chosen because it enforces a consistent formatting rule~\cite{googlestyle} that does not allow custom options, thus ensuring consistent code formatting. 
Subsequently, Antlr~\cite{antlr4} is utilized to extract code blocks from source code files. 
Antlr parses each token with its lexical types, such as ``TypeIdentifier'' and ``VariableDeclarator''. 
This allows us to replace custom names to effectively handle Type-2 variations. 
Instead of replacing different types of tokens indiscriminately, we assign unique dummy names to different token types, recognizing that the diversity of types contains essential information that has a significant impact on the calculation of information entropy. 
The comprehensive replacement table is provided in \Cref{tab:token_replacements}. 
We also preserve the names of invoked functions for two reasons.
Firstly, these names carry unique information and are seldom renamed, particularly those from external modules and libraries. 
Secondly, the distinctiveness of calling a particular function significantly reduces the likelihood of false positives. 
An example of the fully processed code is presented in \Cref{tab:code_transformation}.

\begin{table*}[htbp]
\centering
\footnotesize
\caption{Lexical Analysis and Weight Assignment Example}
\label{tab:code_transformation}
\begin{tabular}{p{0.4\textwidth}cl}
\hline
\textbf{Original Code} & \textbf{Weight} & \textbf{Transformed Code} \\ \hline
public void read(String key, InputStream in) \{ & & \\
\ \ File f = new File(mStoreDir, key); & 0.150 & TypeID LocVar = new TypeID(ID, LocVar); \\
\ \ \begin{tabular}[c]{@{}l@{}}OutputStream os = new BufferedOutputStream\\(new FileOutputStream(f));\end{tabular} & 0.176 & \begin{tabular}[c]{@{}l@{}}TypeID LocVar = new TypeID(new TypeID(LocVar));\end{tabular} \\
\ \ try \{ & - & -\\
\ \ \ \ byte[] data = new byte[1024]; & 0.138 & PmtType[] LocVar = new PmtType[Num]; \\
\ \ \ \ int nbread; & 0.042 & PmtType LocVar; \\
\ \ \ \ while ((nbread = in.read(data)) != -1) \{ & 0.280 & \begin{tabular}[c]{@{}l@{}}while ((LocVar = LocVar.read(LocVar)) != -Num) \{\end{tabular} \\
\ \ \ \ \ \ os.write(data, 0, nbread); & 0.118 & LocVar.write(LocVar, Num, LocVar); \\
\ \ \} & -& -\\
\ \ finally \{ & -& -\\
\ \ \ \ in.close(); & 0.048 & LocVar.close(); \\
\ \ \ \ os.close(); & 0.048 & LocVar.close(); \\
\ \ \} & -& -\\
\} & -& -\\ \hline
\end{tabular}
\end{table*}

\begin{algorithm}
\caption{Weighting}
\label{alg:weight}
\begin{algorithmic}[1]
\Function{Weighting}{$all\_lines$}
    \State $\text{// resolve token scores}$
    \State $\text{valid\_lines} \gets \text{all\_lines.filter\_lines()}$
    \State $\text{TFIDF\_map} \gets \text{calculate\_TFIDF\_map()}$
    \For{$\text{codeline\textbf{ in }valid\_lines}$}
        \State $H_{line} \gets 0.0$
        \For{$\text{token\textbf{ in }codeline.get\_token\_list()}$}
            \State $H_{line} \gets H_{line} + H_{token}(token)$
        \EndFor
    \EndFor
    \State $\text{// weighting}$
    \State $\text{total\_score} \gets 0$
    \For{each $\text{codeline\textbf{ in }valid\_lines}$}
        \State $\text{total\_score} \gets \text{total\_score} + \text{codeline.score}$
    \EndFor
    \For{each $\text{codeline\textbf{ in }valid\_lines}$}
        \If{$\text{total\_score} = 0$}
            \State $\text{codeline.weight} \gets 0$
        \Else
            \State $\text{codeline.weight} \gets \text{codeline.score} / \text{total\_score}$
        \EndIf
    \EndFor
\EndFunction
\end{algorithmic}
\end{algorithm}

\subsection{Weight Assignment}
The primary objective of Weight Assignment is to allocate greater weight to lines that embody more crucial logic. According to the definition in \Cref{sec:essence_def}, we managed to evaluate the importance of code by the amount of information. Aizawa et al.~\cite{AIZAWA200345} mathematically defined the ``probability-weighted amount of information'' (PWI) and demonstrated its calculation by TF-IDF. Inspired by his work, we apply TF-IDF to evaluate the amount of information contained by each line within a given function. TF-IDF is a statistical measure used to evaluate how important a word is to a document in a collection or corpus. The importance increases proportionally to the number of times a word appears in the document, but is negatively affected by the frequency of the word in the whole corpus. In this work, we regard a code block as the whole ``corpus'' and lines as the ``documents''. Then, the information score $H_{token}$ of a particular token $t$ is calculated using the formula below:
$$
H_{token}(t) = \frac{C(t)}{CA_{token}}\times \log\frac{CA_{line}}{CL(t)}
$$
where $C(t)$ is the count of the specific token $t$ within the function, $CA_{token}$ (i.e., count all) denotes the total number of tokens within the function, $CL(t)$ (i.e., count line) is the number of lines containing the token $t$, and $CA_{line}$ is the total count of non-empty lines in the function.
The algorithm of weighting is shown in \Cref{alg:weight}.
Initially, a heuristic filter is applied to remove empty lines or lines containing only control structures, such as \textit{try} or \textit{catch} (e.g., \textit{catch (SomeExceptionOrError e)}) expressions. These lines, primarily serving as fail-safe mechanisms, do not contribute to the core logic. Retaining them could hinder the identification of clone candidates. Following this exclusion, weights are assigned to the remaining, filtered code lines.

The weighting process for a given function starts by calculating the information score $H_{token}(t)$ of all tokens. Then, the amount of information of a particular line $H_{line}(l)$ is determined by aggregating the information scores of its tokens as implemented from lines 5 to 9:
$$
H_{line}(l) = \sum_t H_{token}(t)
$$
Subsequently, from line 9 to line 17, the weight of each line is determined based on the proportion of the information score.  
As depicted in \Cref{tab:code_transformation}, our analysis encompasses a total of eight lines of code, each associated with computed weights. 
Notably, different lines of code exhibit varying weights. 
For instance, the line ``\textit{while ((nbread = in.read(data)) != -1) \{}'' stands out with the highest weight of 0.280 while the weight of line ``\textit{in.close();}'' is only 0.048.

\subsection{Locate \& Filter}\label{sec:locate}
\begin{figure}
    \centering
    \includegraphics[width=0.6\linewidth]{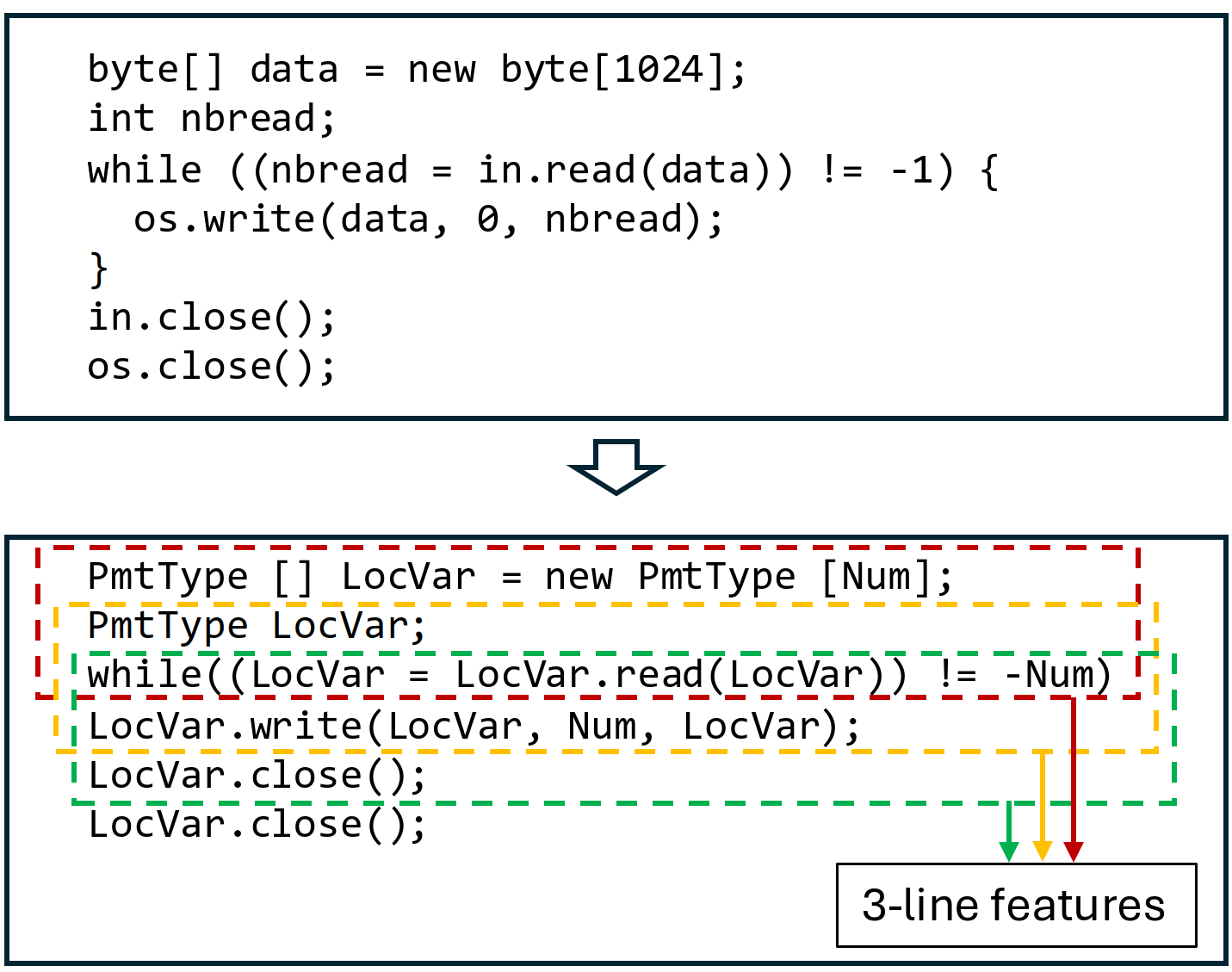}
    \caption{Example of 3-line features }
    \label{fig:nline}
\end{figure}

To identify potential code blocks as clone candidates, we start by building a reverse index, as illustrated in \Cref{alg:reverseindex}. In line 3, a reverse index dictionary is established using hash features as keys, and the corresponding values are the functions sharing these hashes. Subsequently, an N-line hash is generated for each function to serve as its features, with \Cref{fig:nline} providing an example when N equals 3. We apply SHA-256 for hash calculation. For a function containing M lines, $M - N + 1$ features will be produced; if M is less than N, only one feature is created. To locate a clone candidate for a new function, we first compute the new function's features, then search the reverse index to identify potential clones. We observed that some features correlate with thousands of functions, complicating the verification process in later stages. Manual examination reveals that many of these combinations of code lines are routine. For instance, the google-java-format might divide a long parameter list into consecutive lines, each holding a single dummy variable name. These common features are less informative and reduce efficiency, prompting us to filter them using a threshold $\sigma$. To set this threshold, we rank features by the number of associated functions and manually inspect the source code for each feature with a team of three researchers (two juniors and one senior). A feature is discarded if the majority finds it unrepresentative. We conclude that features linked with more than 0.7\% of the total number of functions are typically commonplace and uninformative, and thus filtered. After identifying potential clone candidates, we proceed to verify each pair of functions.

\begin{algorithm}
\caption{Reverse Index}
\label{alg:reverseindex}
\begin{algorithmic}[1]
\Function{BuildReverseIndex}{$\text{func, valid\_lines, N}$}
    \State $\text{// building reverse index}$
    \State $\text{re\_index\_map} \gets \text{empty dictionary}  $
    \If{$\text{valid\_lines.size()} \leq \text{N}$}
        \State $\text{feature} \gets \text{empty string}$
        \For{$\text{codeline\textbf{ in }valid\_lines}$}
            \State $\text{feature} \gets \text{feature} + \text{codeline.get\_hash()}$
        \EndFor
        \State $\text{re\_index\_map.put}\text{(calc\_hash(feature), func)}$
    \Else
        \For{$i \gets 0$ \textbf{to} $\text{valid\_lines.size() - N + 1}$}
            \State $\text{union\_hash} \gets \text{``''}$
            \For{$j \gets 0$ \textbf{to} $\text{N - 1}$}
                \State $\text{union\_hash} \mathrel{{+}{=}} \text{valid\_lines.get}(i+j).\text{get\_hash()}$
            \EndFor
            \State $\text{re\_index\_map.put(calc\_hash(union\_hash), func)}$
        \EndFor
    \EndIf
    
\EndFunction
\end{algorithmic}
\end{algorithm}


\subsection{Verify}\label{sec:verify}
\begin{algorithm}[h]
\caption{Verify the similarity between functions}\label{alg:verify}
\begin{algorithmic}[1]
\Function{verify}{$\text{Func } f1, \text{Func } f2$}
    \State $unique\_hash\_set \gets \text{empty set of strings}$
    \For{$\text{weighted\_line\textbf{\textbf{ in }}} f1$}
        \State $\text{unique\_hash\_set.{add(get\_hash(weighted\_line))}}$
    \EndFor
    \For{$\text{weighted\_line\textbf{ in }} f2$}
        \State $\text{unique\_hash\_set.add(get\_hash(weighted\_line))}$
    \EndFor
    \State $D \gets \text{unique\_hash\_set.get\_length()}$
    \State $\text{build\_hash\_index()}$
    \State ${V\_1} \gets \text{new double[$D$]}$
    \State ${V\_2} \gets \text{new double[$D$]}$
    \For{$\text{code\_line\textbf{ in }} f1$}
        \State $\text{V\_1[index\_of(code\_line)]} \mathrel{{+}{=}} \text{code\_line.weight}$
    \EndFor
    \For{$\text{code\_line\textbf{ in }} f2$}
        \State $\text{V\_2[index\_of(code\_line)]} \mathrel{{+}{=}} \text{code\_line.weight}$
    \EndFor
    \State ${result} \gets \text{cosine\_similarity(V\_1, V\_2)}$
    \State \Return ${result}$
\EndFunction
\end{algorithmic}
\end{algorithm}
\Cref{alg:verify} presents the verification algorithm. Between lines 3 and 6, for each pair of clone candidate functions, we first compute the hash for every weighted line. Then, at line 7, we merge the two sets of hashes to ascertain the count of unique hashes, represented by $D$. Line 8 involves assigning indexes to each unique code line. Consequently, from lines 9 to 14, we depict each function as a $D$-dimensional vector, where each dimension corresponds to a unique hash. The value of a dimension is determined by the weight of the associated code line, with weights of duplicate code lines aggregating in the same dimension. Finally, we compute the cosine similarity between the two $D$-dimensional vectors using the designated formula.
$$
cosine\_similarity = \frac{V_{D1}\cdot V_{D2}}{|V_{D1}|\cdot |V_{D2}|}
$$
where $V_{D1}$ and $V_{D2}$ are the $D$-dimension vectors for the two functions and $|V_{D}|$ is the modular of the vector. After calculating the similarity, we compare it with a predefined similarity threshold, denoted by $\theta$. A clone candidate is considered an actual clone only if the calculated similarity surpasses $\theta$.

\section{Evaluation}
In this section, 
we begin by optimizing the parameters necessary for \toolname. 
Subsequently, we assess the precision and recall of \toolname in detecting essence clones and general clones. 
Lastly, we investigate \toolname's scalability across different input sizes. 
Specifically, we mainly answer the following research questions:
\begin{itemize}
    \item \emph{RQ1: What are the most suitable parameters for \toolname in detecting code clones?} (\Cref{sec:rq1})
    \item \emph{RQ2: What is the performance of \toolname in detecting essence code clones?} (\Cref{sec:rq2})
    \item \emph{RQ3: What is the performance of \toolname in detecting general code clones?} (\Cref{sec:rq3})
    \item \emph{RQ4: What is the runtime overhead of \toolname in detecting code clones?} (\Cref{sec:rq4})
\end{itemize}

\subsection{Experimental Settings}

\subsubsection{Experimental Dataset.} \label{section:dataset}

Our experiments are conducted on the extensively utilized \emph{BigCloneBench} (BCB) dataset \cite{big,svajlenko2014big}, comprising 25,000 open-source projects and 365 million lines of source code. 
The clone pairs within the BCB dataset are categorized at the function level, which aligns seamlessly with the granularity of our detection method. 
Notably, these clone pairs are meticulously labeled with clone types, facilitating the evaluation of \toolname's detection effectiveness across various clone pair classifications. 
It should be noted that \emph{\toolname} focuses on detecting general and essence Type-3 code clones.

Although some recent studies suggest that the BigCloneBench ground truth is flawed, these issues do not affect our evaluation. The critique highlights that a large portion of true positives remain unvalidated and that only a small set of true negatives has been identified, leaving most potential pairs in the dataset with unknown ground truth~\cite{krinke2022bigclonebench}. Such incompleteness could hinder the training of machine learning-based tools, as missing labels might lead to uncertain results during tool training. However, our experiments focus on evaluating recall and precision, meaning that the existing true positives are fully usable for our purposes.


\subsubsection{Comparative Tools.}
Numerous code clone detection tools have been proposed, with many demonstrating commendable performance in terms of both effectiveness and scalability. 
However, it's worth noting that a significant portion of these tools are not open source. 
As a result, for comparative purposes, we have selected the following open-source code clone detection methods to contrast with our approach.
\begin{itemize}
    \item \textbf{SourcererCC} \cite{sajnani2016sourcerercc}  assesses code clone similarity by computing the token overlap between two methods.
    \item \textbf{LVMapper} \cite{lvmapper2020}  identifies clones with significant variances by locating two sequences that are similar yet contain numerous discrepancies.
    \item \textbf{CCAligner} \cite{wang2018ccaligner} detects clones that feature large gaps by utilizing a code window that allows for an edit distance \emph{e}.
    \item \textbf{Siamese} \cite{Ragkhitwetsagul2019SiameseSA} converts Java code into four distinct code representations to identify various clone types.
    \item \textbf{NiCad} \cite{roy2008nicad} employs text-line comparisons and the longest common subsequence method for clone detection.
    \item \textbf{NIL} \cite{nil2021} utilizes N-grams, an inverted index, and the longest common subsequence to identify clones with significant variances.
    \item \textbf{Deckard} \cite{jiang2007deckard} groups characteristic vectors of each abstract syntax tree (AST) subtree using specific rules to identify code clones.
\end{itemize}

Recently, some tools utilize learning-based techniques for clone detection~\cite{dou2024cc2vec,zhang2019astnn}, However, Wang et al.~\cite{wang2023comparison} has tested them and proved that such techniques require a large amount of ground truth data for training; Otherwise they exhibit poor generalizability. Additionally, the training overhead is time-consuming. They also have high hardware requirements, which we currently cannot meet. Thus, we did not compare them in our work.

\subsubsection{Metrics}
We use recall, precision, and F1-score to measure \toolname, which are standardized and align with those used in existing research \cite{sajnani2016sourcerercc, wang2018ccaligner, nil2021}.

\subsection{Parameter Setting}\label{sec:rq1}
\toolname employs two critical parameters: the clone identification similarity threshold, represented by $\theta$, and the N-line size for function features, utilized in pinpointing clone candidates, denoted as $N$. 
We set $\theta$ at 0.7, in accordance with practices observed in numerous studies~\cite{roy2008nicad,nil2021,code2img2023}. 
The configuration of $N$ requires careful consideration.
A lower $N$ can boost recall by flagging potential code clones with fewer line matches, but this might also lead to decreased efficiency and heightened memory usage due to an increase in the volume of clone candidates.
On the other hand, a higher $N$ accelerates scan speed but could diminish the number of identified clones, thus adversely affecting the recall of the tool. 
The objective is to identify an optimal $N$ value that ensures both efficiency and accuracy.
To this end, we conducted experiments to determine the optimal $N$ value within the range of 1 to 5.

\begin{table}[h]
\centering
\caption{ Recall and Time Cost with Various N Values }
\begin{tabular}{cc|ccccc}
\hline
&\textbf{N} & \textbf{1} & \textbf{2} & \textbf{3} & \textbf{4} & \textbf{5} \\ \hline
\multirow{6}{*}{Recall} 
                        &T-1 & 100 & 100 & 100 & 100 & 100 \\
                        &T-2 & 99 & 99 & 99 & 99 & 99 \\
                        &VST3 & 97 & 97 & 95 & 95 & 93 \\
                        &ST3 & 74 & 66 & 66 & 63 & 59 \\
                        &MT3 & 9  & 8 & 8 & 7 & 5 \\
                        &WT3/T4  & 1  & 1 & 1 & 0 & 0\\
\hline
\multicolumn{2}{c|}{Precision} & 69 & 90 & 98 & 98 & 99 \\
\hline
\multicolumn{2}{c|}{Time}  & 124m22s & 29m34s & 9m18s & 7m17s & 5m02s \\ 
\hline
\end{tabular}
\label{table:param_n}
\end{table}

We evaluate the recall and precision using BCB and the results are shown in \Cref{table:param_n}. 
For Type-1, the recall results remain consistently perfect at all values. 
Type-2 and VST3 also demonstrate high effectiveness, with recall slightly decreasing as $N$ increases, suggesting a degradation in performance beyond $N$=3. 
The recall decline is more pronounced in ST3, where the highest score is observed at $N$=1, gradually decreasing with each increase in $N$, further emphasizing the impact of configuration on performance. 
\lida{In contrast to recall, the precision score performs increase as $N$ increases. When $N$ is 1, \toolname reports too much false positives and has a poor precision score of 69\%. The precision improves significantly when changing $N$ from 1 to 2, and this trend becomes slow when $N$ is larger than 3.}
The time efficiency exhibits a significant improvement as $N$ increases, with the execution time dramatically reduced from about 124 minutes at $N$=1 to just around 9 minutes at $N$=3, beyond which the marginal time savings do not compensate for the loss in type performance accuracy. 
These observations collectively lead to the conclusion that $N$=3 represents the optimal setting, striking a balanced trade-off between the execution time and the performance across all types. 


\begin{titleEnv}
\emph{
\textbf{Summary:}
In this paper, the clone identification similarity threshold $\theta$ is 0.7 and the line number for the indexing feature $N$ is 3.
}
\end{titleEnv}

\subsection{General Clone}\label{sec:rq3}
In this part, we assess the effectiveness of \toolname on general clones with a widely used benchmark BCB~\cite{big}.

\subsubsection{Tool Selection}
We select eight state-of-the-art clone detection tools, including CCAligner~\cite{wang2018ccaligner}, SourcererCC~\cite{sajnani2016sourcerercc}, Siames~\cite{Ragkhitwetsagul2019SiameseSA}, NIL~\cite{nil2021}, Nicad~\cite{roy2008nicad}, LVMapper~\cite{lvmapper2020}, and Decard~\cite{jiang2007deckard}. 

\subsubsection{Experiment Setup}
Recall is assessed through BCB ground truth, the ground truth is separated into 6 types thus there are 6 rows of recalls for each tool. However, according to BCB~\cite{big}, the ground truth is not complete, therefore precision is evaluated manually. 
We randomly select 400 clone pairs and manually examine them by researchers, ensuring that each result is verified by at least two researchers. After evaluating the precision, we calculate the F1-score for each tool by clone type, using the overall precision and their corresponding recall. While using the overall precision may introduce some bias, it allows for a clearer understanding of the comparison results.

\begin{table*}
\centering
\caption{Recall and Precision Results for BigCloneBench}
\begin{tabular}{cc|ccccccccc}
\hline
&\textbf{Tool} & \rotatebox{60}{\textbf{\toolname}} & \rotatebox{60}{\textbf{CCAligner}} & \rotatebox{60}{\textbf{SourcererCC}} & \rotatebox{60}{\textbf{Siames}} & \rotatebox{60}{\textbf{NIL}} & \rotatebox{60}{\textbf{Nicad}} & \rotatebox{60}{\textbf{LVMapper}} & \rotatebox{60}{\textbf{Deckard}} \\ \hline
\multirow{6}{*}{Recall} 
&T1 &100	&100	&94	&100	&99	&99	&99	& 60  \\
&T2 &99	&100	&78	&96	&97	&99	&99	& 52  \\
&VST3 &95	&99	&54	&85	&94	&98	&98	& 62  \\
&ST3 &66	&65	&12	&59	&67	&70	&81	& 31  \\
&MT3 &8	&14	&1	&14	&11	&1   &19	& 12  \\
&WT3/T4 &1	&-	&1	&-	&-	&-	&1	& -   \\
\multicolumn{2}{c|}{Overall Recall} & 28	&30	&20	&29	&28	&25	&32	&18 \\
\hline
\multicolumn{2}{c|}{Precision} & 98 & 61 & 100 & 98 & 94 & 80 & 59 & 35  \\ 
\hline
\multicolumn{2}{c|}{Overall F1-Score} & 44	&40	&34	&45	&44	&39	&42	&26 \\
\hline
\end{tabular}
\label{table:bcb_result}
\end{table*}

\subsubsection{Results}
The comprehensive evaluation result of clone detection tools is shown in \Cref{table:bcb_result}. 
\toolname demonstrates remarkable performance across a variety of clone types, underscoring its effectiveness in the field. 
Notably, \toolname achieves a perfect recall of 100\% in T1 clones, indicating its unmatched ability to identify exact clones. 
This performance is consistent with other top-performing tools such as CCAligner and SourcererCC. 
In detecting T2 and VST3 clones, \toolname also exhibits scores closely rivaling or matching the best, highlighting its robustness in detecting more complex, slightly modified clones. 
In contrast, \toolname's recall slightly decreases in ST3 and MT3 clones, particularly in MT3 clones. 
This decrease is attributed to \toolname's focused approach in identifying transplanted core logic, as duplicating non-core logic lines does not qualify as an essence clone. According to Svajlenko et al.~\cite{svajlenko2014big}, MT3 and WT3/T4 clones are function pairs initially selected based on semantic similarity, then further categorized by line-level similarity after full normalization. Nevertheless, our observations show that most MT3 and WT3/T4 actually exhibit syntax similarity primarily in structural code, such as try-catch exception handling blocks, while other syntactic similarities are generally limited to single lines. A single line of code holds little significance for \toolname. \toolname analyzes features in N-lines to locate and filter non-essence clone pairs, intending to provide more robust performance in detecting core logic.  (see \Cref{sec:locate}).
Despite this, \toolname surpasses several other tools in these categories, boasting a superior precision score of 98\% in accurately detecting clones when they are present.
We further calculate the overall recall of all clone types for F1-score calculation. \toolname has a F1-score of 44\% which is slightly worse than Siames with 45\%. Siames utilizes multiple code representations and query reduction to achieve a slightly more accurate detection but requires much more execution time (see \Cref{sec:rq4}).
This data highlights \toolname's strengths and its adeptness at detecting exact and nearly exact clones, establishing it as a frontrunner in general clone detection.


\begin{titleEnv}
\emph{
\textbf{Summary:}
\toolname shows impressive performance across diverse types of clones, highlighting its effectiveness in the field.
}
\end{titleEnv}

\subsection{Essence Clone}\label{sec:rq2}
In this part, we evaluated the accuracy of \toolname in detecting essence clones with precision, recall, and F1-score. 

\subsubsection{Project Selection}
To assess essence clone detection, we selected four Java projects previously used in studies~\cite{wang2018ccaligner,nil2021}, namely Ant, Maven, JDK, and OpenNLP. 
Due to the unavailability of the original versions of some projects, we opted for their latest versions. 
These projects vary in code volume, and their basic characteristics are detailed in \Cref{table:targetsystems}. 

\begin{table}[h]
\centering
\caption{Target Systems}
\begin{tabular}{ccccc}
\hline
\textbf{Project} & \textbf{Version} & \textbf{Files} & \textbf{Code} \\ \hline
Ant &rel-1.10.14 & 1,327  & 146,089 \\
JDK & 23-16 & 49,648  & 5,586,739 \\
Maven & 4.0.0-alpha-13 & 1,391   & 100,529 \\
OpenNLP & 2.3.2 & 1,080  & 77,937 \\ \hline
\end{tabular}
\label{table:targetsystems}
\end{table}

\subsubsection{Tool Selection}
We selected two state-of-the-art tools for comparison: SourcererCC~\cite{sajnani2016sourcerercc} and NIL~\cite{nil2021}. 
SourcererCC is a scalable code clone detection system, which is designed to efficiently identify code clones within large codebases.
We choose this tool due to its frequent comparison in prior studies, highlighting its straightforward methodology. 
NIL, on the other hand, is a clone detection technique tailored for large-variance clones. 
Since it is the most scalable token-based large-variance code clone detector, we also choose it as our comparative tool.

\subsubsection{Experiment Setup}~\label{sec:ec-setup}
We assess the tools using recall, precision, and the F1-Score. Recall measures the proportion of true positives accurately identified. In this experiment, we adopt the definition of recall from Wang et al.~\cite{wang2018ccaligner} as follows:
$$
Recall = \frac{T_p}{\text{Union of }T_p}
$$
where $T_p$ denotes the true positive outcomes identified by a single tool, and the denominator represents the union of true positive results from all three tools. 
Precision assesses the accuracy of positive predictions. 
In this experiment, it is determined by manually validating a random sample of the clones detected by the tools (100 for each project) as previous works done~\cite{nil2021}. 
This process involves three researchers, where two junior researchers initially validate the results independently. 
Discrepancies are resolved by a senior researcher, who makes the final determination. 
After that, we use F1-score that offers a balanced harmonic mean of precision and recall, indicating the overall performance of a predictive model.

\begin{figure*}[htbp]
\begin{minipage}[t]{0.49\textwidth}
\begin{lstlisting}[escapechar=@, language=Java]
protected void dieOnCircularReference(Stack<Object> stk, Project p){
    if (isChecked()) {
        return;
    }
    if (isReference()) {
        super.dieOnCircularReference(stk, p);
    } else {
        @\colorbox{yellow}{for (ResourceSelector resourceSelector : }@
            @\colorbox{yellow}{resourceSelectors) \{}@
            @\colorbox{yellow}{if (resourceSelector instanceof DataType) \{}@
                @\colorbox{yellow}{pushAndInvokeCircularReferenceCheck(}@
                  @\colorbox{yellow}{resourceSelector, stk, p);}@
            @\colorbox{yellow}{\}}@
        @\colorbox{yellow}{\}}@
        @\colorbox{yellow}{setChecked(true);}@
    }
}
\end{lstlisting}
\end{minipage}
\hfill 
\begin{minipage}[t]{0.49\textwidth}
\begin{lstlisting}[escapechar=@, language=Java]
protected void dieOnCircularReference(Stack<Object> stk, Project p){
    if (isChecked()) {
        return;
    }
    if (isReference()) {
        super.dieOnCircularReference(stk, p);
    } else {
        @\colorbox{yellow}{pushAndInvokeCircularReferenceCheck(zips, stk, p);}@
        @\colorbox{yellow}{pushAndInvokeCircularReferenceCheck(tars, stk, p);}@
        @\colorbox{yellow}{setChecked(true);}@
    }
}

\end{lstlisting}
\end{minipage}
\caption{Non-Essence Clone Example from Ant}
\label{fig:not-ec}
\end{figure*}


\subsubsection{Results}\label{sec:ec_result}
\begin{table}[h]
\centering
\caption{Clone Detection Tool Performance}
\begin{tabular}{ccccc}
\hline
\textbf{Project} & \textbf{Tool} & \textbf{Recall} & \textbf{Precision} & \textbf{F1-Score} \\ \hline
\multirow{3}{*}{Ant} & SourcererCC & 49 & 94 & 64 \\
                     & NIL & 68 & 84 & 75 \\
                     & \toolname & 82 & 96 & \textbf{89} \\ \hline
\multirow{3}{*}{Maven} & SourcererCC & 59 & 88 & 71 \\
                     & NIL & 76 & 78 & 77 \\
                     & \toolname & 84 & 88 & \textbf{86} \\ \hline
\multirow{3}{*}{OpenNLP} & SourcererCC & 38 & 85 & 53 \\
                     & NIL & 56 & 76 & 65 \\
                     & \toolname & 85 & 79 & \textbf{82} \\ \hline
\multirow{3}{*}{JDK} & SourcererCC & 40 & 90 & 55 \\
                     & NIL & 54 & 78 & 64 \\
                     & \toolname & 83 & 80 & \textbf{82} \\
\hline
\end{tabular}
\label{table:clone_detection_performance}
\end{table}

\Cref{table:clone_detection_performance} shows the overall results of the three tools on the four projects.
Among the tools, \toolname consistently outperforms SourcererCC and NIL across all projects, indicating its superior ability to accurately identify code clones without compromising on the number of actual clones detected. 
Notably, in the Ant project, \toolname achieves the highest F1-score of 0.885, combining an excellent precision of 0.960 with a strong recall of 0.821. 
This trend is mirrored in other projects, with \toolname maintaining a leading position, particularly exemplified in the OpenNLP project where it scores an F1-score of 0.820. 
However, SourcererCC shows the lowest F1-scores across the board due to its low recall value across all projects, highlighting potential limitations in its clone detection capabilities, especially in projects with complex code bases like OpenNLP and JDK. 
\toolname generally exhibits higher precision than NIL, attributed to its enhanced ability to discern key logical differences between functions. 
For example, as demonstrated in \Cref{fig:not-ec}, two functions from the Ant project illustrate this point.
The primary logic of the left function, spanning from lines 8 to 14, involves circulation checking for a specific type of resource selector, while the function on the right, from lines 8 to 10, performs a manual resource check with hardcoded values.
Despite superficial similarities in their initial lines from 2 to 7, their core logics diverge significantly, rendering them not a pair of essence clones. 
In contrast, NIL tends to recognize them as clones due to their shared token sequences, overlooking their fundamental logical differences. 

Our observations reveal that the same tool's performance varies across different projects. 
In Ant and Maven, the recall gap between \toolname and other tools is markedly smaller than in OpenNLP and JDK. 
A thorough investigation into the nature of the functions within these projects uncovered that functions in Maven and Ant are generally straightforward and concise, making them easier to detect. 
Conversely, functions in JDK and OpenNLP tend to be more complex and lengthier, often containing extensive duplicated code logic, such as processing various types of variables through if/else or switch expressions, where the logic structure is frequently repeated with variable iterations.
In such scenarios, \toolname demonstrates superior capability in identifying and capturing this heavily repeated code logic. 
It does so by aggregating identical logic segments, summing up their weights, and highlighting the significance of these repeated snippets during the essence clone verification process. 
This functionality enables \toolname to effectively pinpoint ``invariant'' code snippets embedded within complex functions, thereby achieving a higher recall rate.



\begin{titleEnv}
\emph{
\textbf{Summary:}
\toolname consistently surpasses both SourcererCC and NIL in performance across various projects, demonstrating its enhanced capability to accurately detect code clones while maintaining the detection of a high number of true clones.
}
\end{titleEnv}


\subsection{Scalability}\label{sec:rq4}
\begin{table*}[ht]
\centering
\caption{Scalability Results}
\label{tab:scalability}
\begin{tabular}{ccccccccc}
\toprule
\textbf{Tool}   & \rotatebox{60}{\textbf{\toolname}}    & \rotatebox{60}{\textbf{CCAligner}}    & \rotatebox{60}{\textbf{SourcererCC}}  & \rotatebox{60}{\textbf{Siames}}  & \rotatebox{60}{\textbf{NIL}}  & \rotatebox{60}{\textbf{Nicad}}    & \rotatebox{60}{\textbf{LVMapper}} & \rotatebox{60}{\textbf{Decard}}\\
\midrule
1K              & 1s                    & 1s                    & 3s                    & 4s               & 1s            & 1s                & 1s      & 1s                \\
10K             & 1s                    & 2s                    & 5s                    & 14s              & 1s            & 3s                & 1s      & 4s               \\
100K            & 20s                   & 6s                    & 7s                    & 45s              & 1s            & 36s               & 4s      & 32s           \\
1M              & 3m46s                 & 11m52s                & 37s                   & 45m1s            & 10s           & 6m13s             & 34s     & 27m12s   \\
10M             & 18m56s                & 29m48s                & 12m21s                & 14h11m           & 1m38s         & 2h10m             & 22m10s  & error   \\
\bottomrule
\end{tabular}
\end{table*}
We evaluate the scalability of \toolname on codebases with multiple sizes and compare the execution time. The dataset we choose is IJaDataset~\cite{ijadataset}, which is a large cross-project dataset in Java that has also been used in many previous studies. The data size varies from one thousand (1K) to ten million (10M), as measured by CLOC~\cite{cloc}. The experiments are run on a quad-core CPU with 12GB of memory, as in the previous studies~\cite{nil2021,wang2018ccaligner,code2img2023}.

The results, as shown in \Cref{tab:scalability}, show that \toolname maintains consistent performance up to a dataset size of 1M, delivering competitive execution times compared to other tools. However, \toolname shows a significant increase in execution time as the dataset size increases, particularly at the 10M level, where its performance begins to lag behind tools such as NIL. NIL manages to keep execution times below 1 minute for datasets up to 100K.

This increase in execution time for \toolname is due to its thorough analysis approach, which includes formatting and variable substitution. While these steps improve \toolname's ability to detect essence clones by providing a more detailed analysis, they introduce a significant overhead, which is particularly noticeable as the dataset size increases. This approach contrasts with tools that skip such exhaustive preprocessing, thereby maintaining shorter execution times across a range of dataset sizes. However, these tools may not be as effective in identifying Type-2 verified clones and essence clones.


\begin{titleEnv}
\emph{
\textbf{Summary:}
\toolname possesses moderate to high Scalability. As the size of the dataset grows, the execution time also increases significantly due to the comprehensive lexical analysis and formatting operations it performs.
}
\end{titleEnv}

\section{Discussion}

\subsection{Threats to Validity}
A significant threat to the validity of this research concerns the measurement of speed in scalability tests. To mitigate memory limitations, we opted to store all related information, including the inverse index, weight information, line hashes, and similarity results, in MongoDB. Numerous performance optimizations, such as batch operations, were implemented to expedite the data query process. Nonetheless, the system architecture still has an impact on the overall execution efficiency. In particular, the construction of the database index and the I/O speeds can markedly affect performance.


\subsection{Limitation}
\lida{
\toolname may be ineffective when the code block has no core logic, for example, when all lines are value settings (e.g., ``this.setFoo(Bar)''), ECScan does not perform well. Also, if the proportion of core logic is too little compared to the whole code block, ECScan may also miss the case.
}


\section{Related Work}
 Current methods for code clone detection are broadly classified into different categories, such as text-based, token-based, tree-based, graph-based, and metric-based tools. 
 These diverse approaches utilize a range of techniques to discern code clones, thereby contributing significantly to the progression of code clone detection.

For text-based clone detection techniques \cite{johnson1994substring, ducasse1999language, roy2008nicad, ragkh2017using, kim2018software, jadon2016code, yu2017detecting, kim2017vuddy}, they compare the similarity of two code snippets by treating the source code as a sequence of lines or strings. 
For instance, Johnson et al. \cite{johnson1994substring} utilize the fingerprint matching method to detect clones, while Ducasse et al. \cite{ducasse1999language} apply string matching methods, treating lines of code as strings to estimate similarity. 
Similarly, Roy et al. \cite{roy2008nicad} detect potential clones using an algorithm based on the longest common subsequence for matching code text. 
These methods focus solely on string matching and do not consider the logic of the code, making them ineffective in identifying complex Type-3 clones.

For token-based clone detection techniques \cite{gode2009incremental, kamiya2002ccfinder, sajnani2016sourcerercc, li2017cclearner, wang2018ccaligner, golubev2021multi, hung2020cppcd}, they apply lexical analysis to convert the source code into token sequences and then detect code clones by scanning them. 
For example, \emph{CCFinder} \cite{kamiya2002ccfinder} extracts token sequences from program code and utilizes transformation rules to enable the detection of Type-1 and Type-2 clones. 
\emph{CCAligner} \cite{wang2018ccaligner} excels in detecting clones with large gaps by employing innovative electron mismatch indices and asymmetric similarity coefficients. 
Golubev et al. \cite{golubev2021multi} propose a modified token bag-based clone detection technique with a multi-threshold search, optimizing clone detection by reducing overlap and effectively identifying Type-3 clones. 

For tree-based clone detection techniques \cite{jiang2007deckard, wei2017cdlh, zhang2019astnn, liang2021astpath, jo2021twopass, pati2017comparison, chodarev2015haskell, hu2022treecen, wu2022amain}, they parse the program code into a parse tree or abstract syntax tree (AST) and then identify clones by performing tree matching. 
For example, \emph{Deckard} \cite{jiang2007deckard} utilizes locality sensitive hashing (LSH) to detect clones by clustering similar vectors computed from AST of. 
\emph{ASTNN} \cite{zhang2019astnn} segments the AST into subtrees based on predefined rules and encodes each subtree independently using a bidirectional recurrent neural network (RNN) model.
\emph{Amain} \cite{wu2022amain} simplifies complex tree analysis using a Markov chains model for code representation, followed by training a classifier with traditional machine learning algorithms.
However, they frequently encounter substantial time overhead and scalability challenges, potentially restricting their effectiveness in handling large-scale codebases.

For graph-based clone detection techniques \cite{krinke2001duplix, komondoor2001pdgdup, wang2017ccsharp, zhao2018deepsim, wu2020scdetector, zou2020ccgraph}, they use program dependency graph (PDG) and control flow graph (CFG) to represent the program code. 
For example, Krinke et al. \cite{krinke2001duplix} and Komondoor et al. \cite{komondoor2001pdgdup} utilize subgraph matching for clone detection. 
However, subgraph matching approaches often entail significant time overhead. 
To mitigate this, \emph{CCSharp} \cite{wang2017ccsharp} introduces two strategies for reducing this overhead. 
The first method involves adjusting the graph structure, while the second method entails filtering the feature vectors. 
In \emph{Deepsim} \cite{zhao2018deepsim}, code control flow and data flow are translated into semantic high-dimensional sparse matrices. These matrices are then transformed into binary feature vectors and fed into a deep learning model to evaluate the functional similarity of the code fragments.
However, like tree-based approaches, the graph-based approach faces scalability challenges due to issues with graph isomorphism and graph matching.


\lida{Text-based and token-based methods offer quick clone detection but lack robustness when names are replaced, limiting their effectiveness in detecting complex code clones. In contrast, graph-based and tree-based methods are adept at identifying complex clones but often require compilation, which can be problematic for small code fragments. Consequently, none of these methods assign weights to differentiate the importance of code lines within a code block, allowing trivial supporting code lines to disproportionately affect the matching results.}

\section{Conclusion}
In this work, we introduce a new category of Type-3 clones, termed essence clones, which focuses on the most critical parts of the code. We developed an innovative tool, \toolname, designed to enhance the detection of essence clones by assigning weights to lines of code based on their information content. Experiment results indicate that \toolname more effectively identifies essence clones compared to existing methods and performs robustly against top clone detectors in detecting general Type-1, 2, and 3 clones.

\section{Data Availability}
The implementation of \toolname can be found on \url{https://anonymous.4open.science/r/ECScan}